\documentclass[review]{elsarticle}

\usepackage{lineno,hyperref}
\usepackage{amsmath}
\usepackage{adjustbox}
\usepackage{color}
\modulolinenumbers[5]

\journal{Engineering Applications of Artificial Intelligence}

\biboptions{authoryear}

\begin{document}

\begin{frontmatter}

\title{An Attention Long Short-Term Memory based system for automatic classification of speech intelligibility}

%\author{Miguel Fern\'andez-D\'iaz, Ascensi\'on Gallardo-Antol\'in\corref{mycorrespondingauthor}}
\author{Miguel Fern\'andez-D\'iaz}
\author{Ascensi\'on Gallardo-Antol\'in\corref{mycorrespondingauthor}}
\address{Department of Signal Theory and Communications, Universidad Carlos III de Madrid}
\address{Avda. de la Universidad, 30, 28911 Legan\'es, Madrid, Spain}

\cortext[mycorrespondingauthor]{Corresponding author}
\ead{gallardo@tsc.uc3m.es}

\begin{abstract}
\textcolor{black}{Speech intelligibility can be degraded due to multiple factors, such as noisy environments, technical difficulties or biological conditions. This work is focused on the development of an automatic non-intrusive system for predicting the speech intelligibility level in this latter case. The main contribution of our research on this topic is the use of Long Short-Term Memory (LSTM) networks with log-mel spectrograms as input features for this purpose. In addition, this LSTM-based system is further enhanced by the incorporation of a simple attention mechanism that is able to determine the more relevant frames to this task. The proposed models are evaluated with the UA-Speech database that contains dysarthric speech with different degrees of severity. Results show that the attention LSTM architecture outperforms both, a reference Support Vector Machine (SVM)-based system with hand-crafted features and a LSTM-based system with Mean-Pooling.}
\end{abstract}


\end{frontmatter}

%\linenumbers

\section{Introduction}
\label{sec:introduction}

Speech represents, for the majority of the people in the world, the most used method of communication. Unfortunately, speech can be degraded because of several factors, including hearing loss, unwanted background noise, malfunctioning devices or even biological conditions that produce impairments in the human speech production system. Speech intelligibility is a measure of the comprehensibility of the speech under certain conditions, such as the ones that can worsen the communication stated above.

This paper deals with the automatic estimation of the speech intelligibility level when speech degradation is due to physiological factors, i.e., in the case of pathological voices. In particular, \textcolor{black}{this paper is focused on dysarthria} \citep{Doyle1997} that is a motor speech disorder where the muscles related to the articulation of phonemes cannot be fully controlled producing a set of symptoms, such as alterations in the speed when talking, choppy speech and involuntary repetitions of phonemes, exaggerated changes in volume and pitch, nasal voice, etc. This malfunctioning can be due to factors as tumors, brain injuries, thrombotic or embolic strokes, or degenerative diseases like Parkinson's Disease (PD), or Amyotrophic Lateral Schlerosis (ALS). Any case, dysarthria usually hinders the communication for the people that suffer from it and can bring some psychological damage to the patients due to this inability to expressing properly.

\textcolor{black}{In this context, the measurement of the speech intelligibility level is useful for a variety of purposes, such as the monitoring of patients following a certain speech therapy or medical treatment.}
%This is, for example, the case of Parkinson's disease where the Unified Parkinson's Disease Rating Scale (UPDRS), which is commonly used to track the progression of this illness, contains two assessment criteria related to the speech communication skills of the patient \citep{Goetz2003}.
In these cases, the gold standard for determining the intelligibility level is the realization of a series of tests where the patients utter words and/or combinations of sounds. Later, a subjective evaluation carried out by one or several specialists in which they listen and personally assign the intelligibility scores, is what determines the comprehensibility of the patients' speech. 

Automatizing this task can be beneficial for several reasons. Firstly, it leaves more time to doctors that could be used for taking care of other patients or other related activities. Secondly, the clinicians' criteria to understand some words are relatively subjective. In fact, as they mainly trust on their hearing skills and might be familiarized with disordered speech, they might assign lower of higher values than the intelligibility really is \citep{Landa2014}. Finally, it provides the specialists with a repeatable and objective assessment. 

In order to avoid these subjectivity issues, the main goal of this work is the development of an objective, automatic and non-intrusive system that can predict the intelligibility level (low, \textcolor{black}{medium, or high}) of a person with dysarthria by analyzing his/her voice. \textcolor{black}{This system is based on the Deep Learning (DL) paradigm, in particular on \emph{Long Short-Term Memory (LSTM)} networks \citep{Hochreiter1997, Gers2003}, that, despite of their success in other speech related tasks,
%such as Automatic Speech Recognition (ASR) \citep{Rao2015}, Speech Emotion Recognition (SER) \citep{Mirsamadi2017} or Cognitive Load (CL) classification from speech \citep{Gallardo-Antolin2019a, Gallardo-Antolin2019b},
to the best of our knowledge, have not been previously used for intelligibility prediction.} In addition, LSTMs are combined with an \emph{attention mechanism} that is able to model the contribution of each temporal frame to the final decision, improving the performance of the proposed system.

The rest of the document is organized as follows: Section \ref{sec:related_work} introduces some related work; Section \ref{sec:speech_intelligibility_classification_system} contains the general description of the intelligibility classification system, as well as detailed information about the features and classifiers used in this work, including our proposal based on attention LSTM models; the database, \textcolor{black}{experiments, and results} are described in Section \ref{sec:experiments}; and Section \ref{sec:conclusions} finishes the document with some conclusions and possible future guidelines.

\section{Related Work}
\label{sec:related_work}

Previous works related to the automatic prediction of pathological speech intelligibility can be divided into two main lines of research \citep{Janbakhshi2019}: \textit{intrusive} or \textit{non-blind} methods that make use of reference speech signals, i.e. intelligible speech uttered by people that do not show difficulties in the speech production; and \textit{non-intrusive} or \textit{blind} approaches that only rely on the speech signals whose intelligibility is going to be measured.

Intrusive or non-blind techniques usually aim to the building of a healthy (intelligible) reference model that can be based, among others, on Gaussian Mixture Models \citep{Bocklet2011}, iVectors \citep{Martinez2015} or spectral bases in the octave band domain \citep{Janbakhshi2019}. The intelligibility measurement is obtained by means of the comparison of the pathological utterance to be assessed to this reference model. Other approaches are based on the assumption that pathological speech should decrease the performance of an automatic speech recognizer trained with healthy voice \citep{Zlotnik2015}. This way, features related to the ASR output, as for example, the word error rate, are used for intelligibility assessment. The main disadvantage that non-blind techniques have, is the need of large amounts of phonetically-balanced healthy data, although recent works have somehow alleviated this problem \citep{Janbakhshi2019}.

Non-intrusive or blind methods typically include the extraction of several sets of hand-crafted features with the optional application of dimensionality reduction techniques over them \citep{Falk2012, Jollife2016, Tharwat2017}, and the use of some machine learning-based regression or classification algorithms \citep{SarriaPaja2012}, such as Support Vector Machines (SVM) \citep{Khan2014} or Random Forests \citep{Byeon2018}. \textcolor{black}{This approach is followed in this work}.

To the best of our knowledge, there is practically no information in the literature on non-blind and blind approaches for pathological speech intelligibility assessment based on deep learning, although, this family of techniques has begun to be employed in other tasks related to disordered speech, such as the automatic detection of Parkinson disease \citep{Upadhya2018} or the identification of dysarthric speakers \citep{Farhadipour2018}.

\textcolor{black}{The dramatic improvements achieved by DL-based techniques in other speech-related tasks, such as Automatic Speech Recognition (ASR) \citep{Rao2015}, Speech Emotion Recognition (SER) \citep{Mirsamadi2017} or Cognitive Load (CL) classification from speech \citep{Gallardo-Antolin2019a, Gallardo-Antolin2019b}, have motivated us to propose their use for intelligibility prediction.} In particular, Long Short-Term Memory networks \citep{Gers2003} are specially suitable for this purpose due to their capability for modeling temporal sequences, as it is the case of speech signals. \textcolor{black}{Moreover, these networks can be combined with the so-called attention models \citep{Chorowski2015, Huang2016, Mirsamadi2017, Gallardo-Antolin2019b}}, that try to learn the structure of the temporal sequences by modeling the relevance of each frame to the task under consideration.
%As we mentioned before, recently, these models have been successfully utilized for, among others, ASR \citep{Chorowski2015}, SER \citep{Huang2016, Mirsamadi2017} or speech CL classification \citep{Gallardo-Antolin2019a, Gallardo-Antolin2019b}.

\textcolor{black}{In summary, the main contribution of this paper is the proposal of an automatic and non-intrusive system for intelligibility level classification of pathological speech that does not require healthy speech data and is based on LSTM networks with log-mel spectrograms as input. In addition, this system is improved by the incorporation of a simple attention model with what further improvements are achieved. Results show that both alternatives outperform a conventional blind system based on hand-crafted features and a SVM classifier.}

\section{Speech intelligibility classification system}
\label{sec:speech_intelligibility_classification_system}

\begin{figure}[t]
	\centering
	\includegraphics[width=1.0\textwidth]{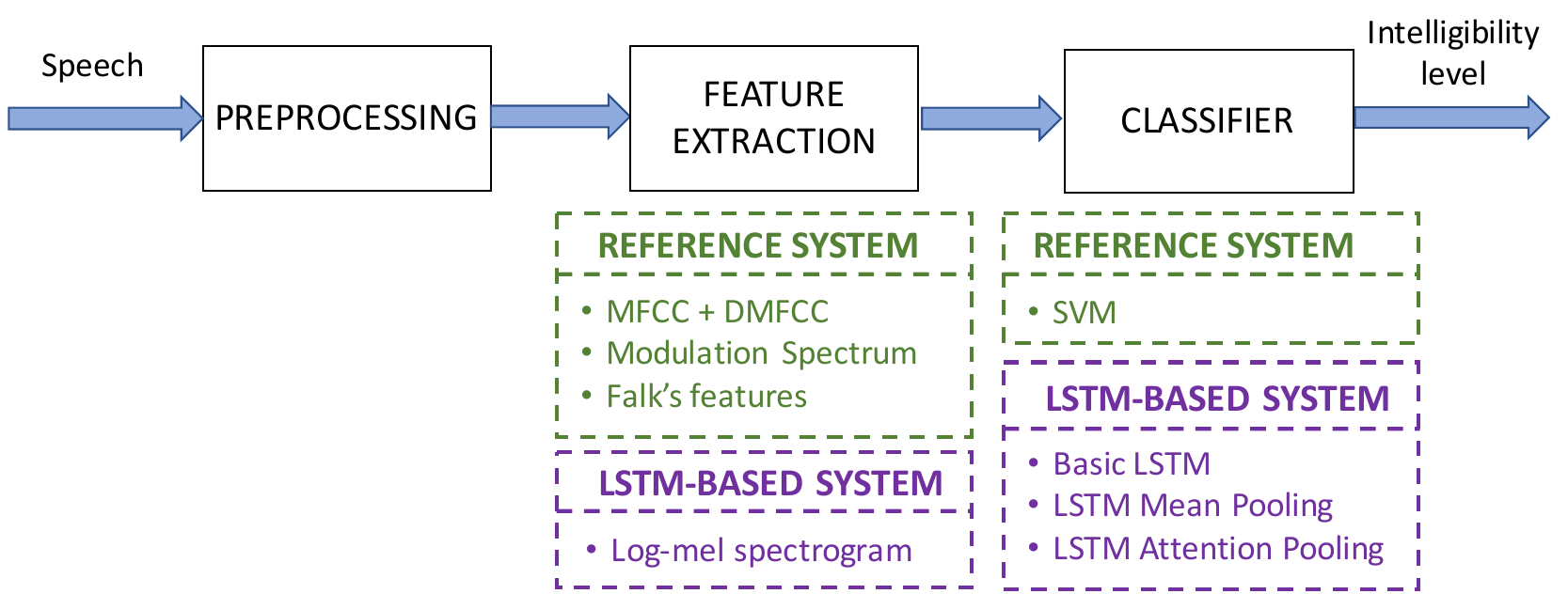}
	\caption{Block diagram of the intelligibility level classification systems developed in this work.}
\label{fig:blocks}
\end{figure}

This Section describes the two kind of systems developed throughout this work with the purpose of classifying the speakers' intelligibility into three categories: low, \textcolor{black}{medium, and high}. On the one hand, the first type of systems is used as reference and consists of the extraction of different sets of hand-crafted acoustic features and a SVM as classifier. On the other hand, the second kind of systems, which comprises our proposal for this task, makes use of a logarithmic scaled mel spectrogram as input and LSTM networks to perform the classification. Both approaches follow a similar sequence of steps, and in order to have a clearer vision of the work, Figure \ref{fig:blocks} depicts a block diagram containing these stages.

\subsection{Preprocessing}
\label{subsec:preprocessing}

The preprocessing step consists of the application of a Voice Activity Detector (VAD) to the raw audio signals in order to remove the silence/noise frames. In particular, the used VAD is the one proposed in \citep{Sohn1999}. The rationale behind this stage is that, in theory, non-speech frames do not convey information about the intelligibility level. However, as it will be shown in Section \ref{sec:experiments}, the removal of the non-speech regions negatively affects the performance of the systems. For this reason, experiments have been carried out in both situations: with and without VAD.

\subsection{Feature extraction}
\label{subsec:feature_extraction}

For the reference system, three different sets of acoustic features have been extracted: (i) Mel-Frequency Cepstrum Coefficients (MFCC) and their first derivatives; (ii) the average energy of the modulation spectrum; and (iii) the set of features proposed in \citep{Falk2012}. For the LSTM-based system, the log-mel spectrograms were used as acoustic characteristics. In the following paragraphs, \textcolor{black}{all of them are briefly described}.

\subsubsection{MFCCs and their first derivatives}
\label{subsubsec:mfcc}

MFCC \citep{Davis1980} is the most popular feature extraction procedure in automatic speech and speaker recognition and also in audio classification tasks (see, for example, \citep{LudenaChoez2016, Abdalmalak2018}). For this reason, these parameters have been tried for the task under consideration. MFCCs are extracted on a frame-by-frame basis by applying the Discrete Cosine Transform (DCT) on the log-mel spectrogram of the speech signal (see Subsection \ref{subsubsec:log_mel_spectrogram}). Once MFCCs are computed, their first derivatives ($\Delta$MFCC) are added to the final acoustic vectors.

\subsubsection{Average energy of the modulation spectrum}
\label{subsubsec:modulation_spectrum}

This set of features are derived from the modulation spectrum of the speech signal \citep{Greenberg1997, VicentePena2006} that measures its long-term temporal dynamics at different modulation frequencies. According to \citep{Falk2012}, the modulation spectrum contains information about several phenomena that can be present in pathological speech, such as, non-habitual intensity and speed variations, imprecise coarticulations or interruptions and disfluencies. 

\begin{figure}[t]
	\centering
	\includegraphics[width=1.0\textwidth]{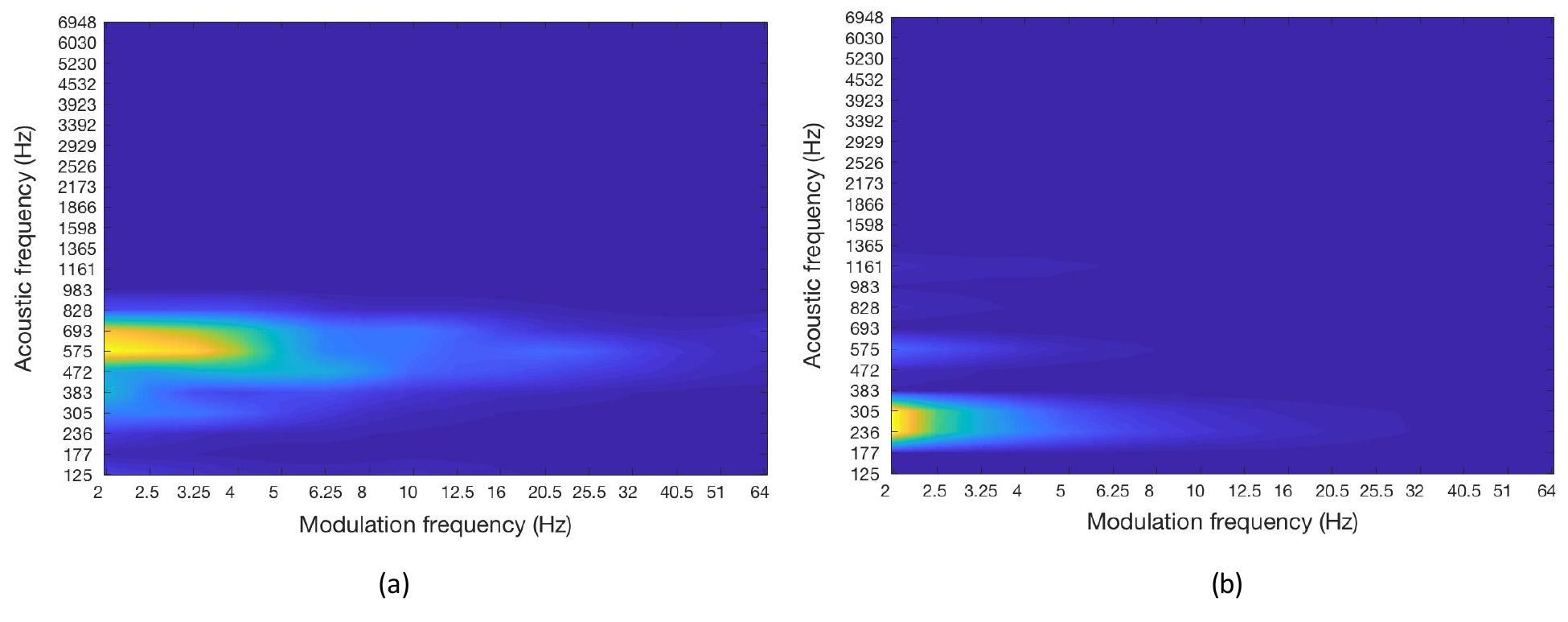}
	\caption{Average energy of the modulation spectrum of a speech recording with (a) high intelligibility  and (b) low intelligibility. Both utterances correspond to the word ``jowls".}
\label{fig:modulation_spectrum}
\end{figure}

The modulation spectrum is computed from a spectral-temporal representation of the audio signal by using the method proposed in \citep{Falk2010}, where the temporal envelopes corresponding to each acoustic frequency band are filtered with a certain modulation filterbank, obtaining the so-called modulation energies. The final set of features consists of the average of these energies over all speech frames. Figure \ref{fig:modulation_spectrum} show two examples of the average of the modulation energies in two different speech recordings, where the horizontal and vertical axes represent, respectively, the modulation and acoustic frequencies. It can been observed that for pathological speakers, the modulation energy is usually very concentrated in low modulation frequencies, as it is the case of the example in Figure \ref{fig:modulation_spectrum} (b), whereas for high intelligibility speakers, the modulation energy spreads over a wider region of frequencies, as in the case of Figure \ref{fig:modulation_spectrum} (a). 

\subsubsection{Falk's features}
\label{subsubsec:falk_features}

This set of acoustic characteristics was initially proposed in \citep{Falk2012} and \citep{SarriaPaja2012} for intelligibility level prediction. It consists of the following six features:

\begin{itemize}

\item Standard deviation of the first derivative of the zeroth order MFCC. This parameter is related to the log-energy of the signal, and can be used to detect anomalies in the speech intensity.

\item Kurtosis of the linear prediction residual. This feature can give information about hoarseness, loss of \textcolor{black}{volume, or breathiness} of the voice.

\item Low to High Modulation Ratio (LHMR). This parameter is a summarization of the information contained in the modulation spectrum of the speech signal (see Subsection \ref{subsubsec:modulation_spectrum}). In particular, it is a quotient that compares the modulation spectral energy at low (lesser than 4 Hz) and high (greater than 4 Hz) modulation frequencies.

\item Three prosody-related features: percentage of voiced segments in the utterance, and the standard \textcolor{black}{deviation, and the} range of the fundamental frequency. The first one can provide information about anomalies in the pronunciation of voiced sounds due to impairments in the phonation organs. The second and third parameters are useful to detect monotone speech (which is a symptom of dysarthria) and shakiness and quiver in the voice.

\end{itemize}

\subsubsection{Log-mel spectrograms}
\label{subsubsec:log_mel_spectrogram}

The last set of features corresponds to the spectrogram of the audio signal that has been first mapped to the mel-frequency spacing \citep{Mermelstein1976} by using an auditory filter bank composed of mel-scaled filters, and later converted to a logarithmic scale. The mel scale is a frequency warping that tries to mimic the non-equal sensitivity of the human hearing at different frequencies.

\subsection{Classifiers}
\label{subsec:classifiers}

%The classification step has been carried out by using two different approaches: the reference system is based on SVMs and our proposal is based on LSTM networks.

\textcolor{black}{As mentioned before, the classification step has been carried out by using two different approaches: SVM and LSTM networks.}

\subsubsection{Support Vector Machines} 
%SVM \citep{Vapnik1964} is one of the most used classifiers, generally combined with hand-crafted features as input. This algorithm discriminates between the different classes by using a set of hyperplanes that satisfy the maximum separation criterion. Although they were initially conceived as binary classifiers, these supervised learning models can be extended to multiclass classification, which in this case has been done by utilizing a one-vs-all strategy.

\textcolor{black}{For the reference system, a SVM \citep{Vapnik1964} with a one-vs-all strategy and a Gaussian kernel has been used.}

\subsubsection{Long Short-Term Memory networks}
SVM-based classifiers are not able to properly model the temporal behavior of sequences. However, the speech signals dynamics play a crucial role for the task considered in this work, as intelligibility is related to events that can be observed in the temporal domain, such as disfluencies, speech interruptions or deficient coarticulations \citep{Doyle1997, Falk2012}. \textcolor{black}{For this reason, and taking into account the success achieved by LSTMs in other speech-related problems, in this paper it is proposed the use of this type of networks for speech intelligibility classification.}

Long Short-Term Memory networks are a special kind of Recurrent Neural Networks (RNNs) that have the ability to store information from the past in the so-called memory blocks \citep{Gers2003}, allowing the information to persist through time. This way, they are able to learn and infer long-term dependencies, overcoming the vanishing gradient problem. Therefore, LSTM outputs depend on the present and previous inputs, and, for this reason, they are very suitable for modeling temporal sequences, as speech.

LSTMs carry out a sequence-to-sequence learning that can be thought as a transformation of an input sequence of length $T$, $x = \{x_1, ..., x_T\}$ into an output sequence $y = \{y_1, ..., y_T\}$ of the same length, assuming that the classification process is easier in the $y$-space than in the $x$-space. However, speech intelligibility classification can be seen as a many-to-one sequence-to-sequence learning problem \citep{Huang2016} as the input is a sequence of acoustic vectors and the final output must be the predicted intelligibility level for the whole utterance (a single value). This means that the information contained in the temporal LSTM output sequence should be compacted into an utterance-level representation $z$, what can be accomplished in different ways. \textcolor{black}{In this paper, it is followed the approach presented by \citep{Mirsamadi2017} for speech emotion recognition, and it is proposed three different LSTM architectures for the task under consideration}, as explained in \textcolor{black}{the} following paragraphs. 

\paragraph{Basic LSTM}
Basic LSTM corresponds to the conventional approach where only the last frame of the LSTM output is passed through the following dense layer. This is equivalent to consider that the utterance-level representation $z$ is computed as $z = y_T$. This strategy assumes that, as in LSTM networks every output relies on previous and present inputs, it could be expected that the last output is the most reliable one since for its computation, the LSTM uses to some extent information from the whole utterance \citep{Zazo2016}.

\begin{figure}[t]
	\centering
	\includegraphics[width=0.6\textwidth]{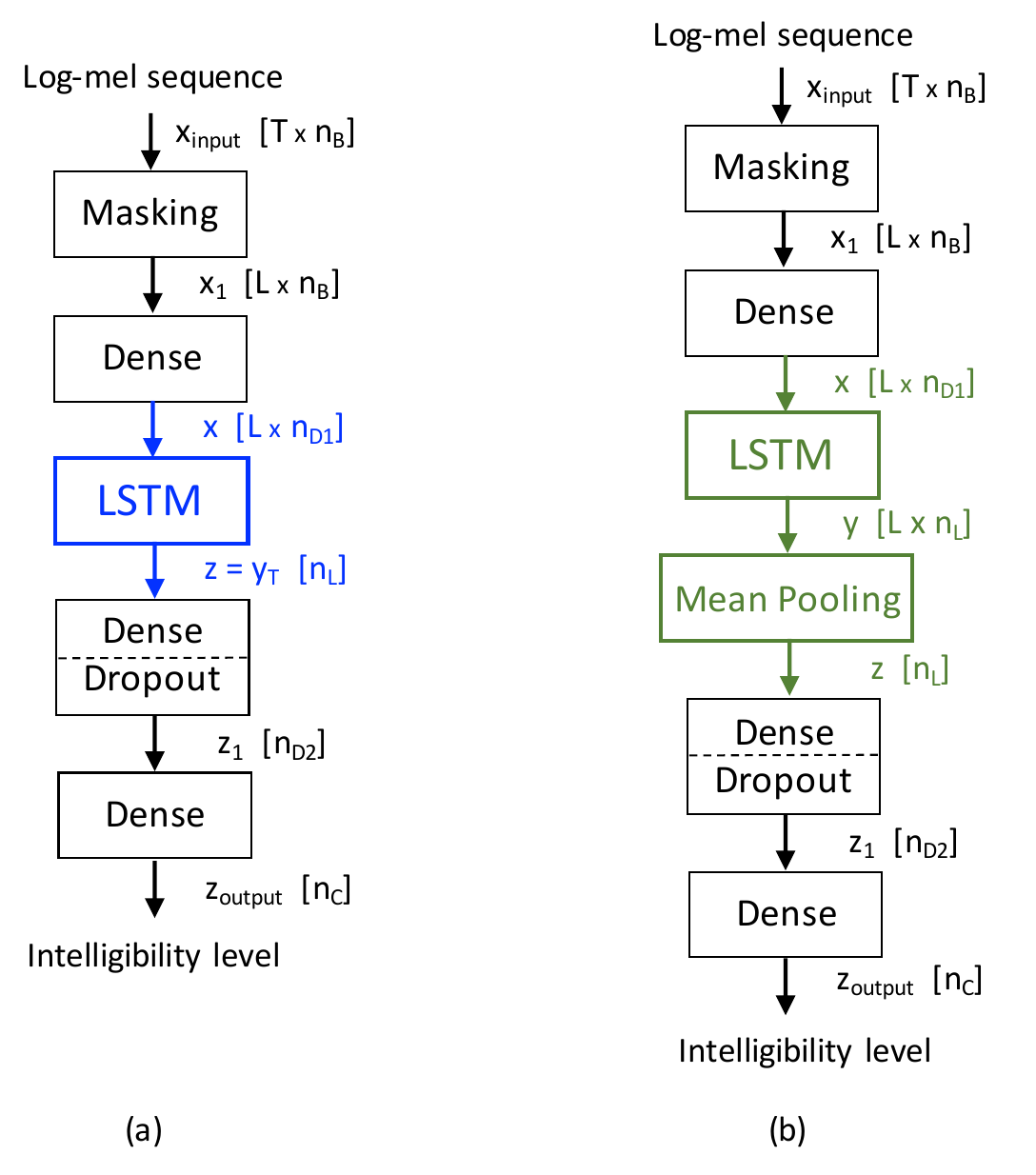}
	\caption{Two different LSTM-based architectures for speech intelligibility classification. (a) Basic LSTM; (b) LSTM with Mean-Pooling. In brackets, the dimension of each variable, where $T$, $n_B$, $L$, $n_{D1}$, $n_L$, $n_{D2}$ and $n_C$, stand for the number of frames of the input signal, the number of mel filters, the length of the LSTM input/output sequence,  the number of neurons in the first dense layer, the number of LSTM units, the number of neurons in the second dense layer and the number of classes (intelligibility levels), respectively.}
\label{fig:lstm_architecture}
\end{figure}

Figure \ref{fig:lstm_architecture} (a) shows the architecture of this basic LSTM-based system. It uses log-mel spectrograms as input features whose dimensions are $T \times n_B$, being $T$ the number of frames of the input signal and $n_B$ the number of mel filters. As log-mel spectrograms do not have the same size for all the speech recordings and, however, the length of the LSTM input sequences need to be set to a fixed value $L$, shorter utterances than this amount are previously padded with dummy values. The Masking layer makes that these masked values not to be used in further computations. Longer utterances are cut if necessary (this only occurs in a few cases as it is shown in Section \ref{sec:experiments}).

Later, a dense layer of $n_{D1}$ neurons where all of them are connected is introduced and the output of this layer goes to a LSTM layer of $n_L$ cells, where, as mentioned before, only the last value of the resulting output LSTM sequences is retained. The output of the LSTM layer goes then into a second dense layer of $n_{D2}$ neurons with dropout in order to avoid overfitting. In the end, a final dense layer with $n_C$ nodes activated by a softmax function performs the multiclass classification. In this case, $n_C$ \textcolor{black}{must match} the number of possible intelligibility levels to be predicted.

\paragraph{LSTM with Mean-Pooling}
\textcolor{black}{The basic LSTM model only returns the last value of the output LSTM sequence, and therefore, it might lead to some information loss,} that can be alleviated by making that every LSTM frame to somehow contribute to the final utterance-level representation $z$. In the case of the Mean-Pooling scheme, $z$ is computed as the average of the output LSTM frames across the whole utterance, as expressed in Equation \eqref{eq:mean_pooling}.

\begin{equation} 
z = \frac{1}{T} \sum_{t=1}^{T} y_t,
\label{eq:mean_pooling}
\end{equation}

Note that Equation \eqref{eq:mean_pooling} can be interpreted as that all LSTM frames contribute equally to $z$ \textcolor{black}{with a weight that is equal to $\frac{1}{T}$}.
%In other words, all frames have the same weight that is equal to $\frac{1}{T}$.

As can be observed in the graphical representation of the LSTM with Mean-Pooling system shown in Figure \ref{fig:lstm_architecture} (b), \textcolor{black}{its} architecture is identical to the basic LSTM one, apart from the Mean-Pooling layer (depicted in green lines). 

\paragraph{LSTM with Attention-Pooling}
As mentioned before, in the LSTM Mean-Pooling model it is assumed that all the LSTM frames reflect the intelligibility level with the same intensity. However, it is reasonable to expect that, within an utterance, certain frames contain more information cues about speech intelligibility than others. In other words, the contribution of each frame should be weighted according to its relevance to the task, enhancing the parts of speech that are more significant, and attenuating or even eliminating the contribution of segments that do not contain useful information.

One of the most recent approaches to address this issue is the use of an attention mechanism, where the utterance-level representation $z$ is computed following the expression in Equation \eqref{eq:attention_pooling},

\begin{equation} 
z = \sum_{t=1}^{T} \alpha_{t} y_t,
\label{eq:attention_pooling}
\end{equation}

In this formulation, $\alpha_{t}$ represents the weight corresponding to the $t$-th LSTM frame. When enough training data is available, it is possible to design complex attention models for the computation of these weights, as, for example, the one described in \citep{Huang2016}. However, since in our case the amount of training data is scarce, the strategy originally proposed in \citep{Mirsamadi2017} for emotion recognition is adopted and is applied to speech intelligibility level classification. In this method, the attention weights are calculated as indicated in Equation \eqref{eq:attention_weights}.

\begin{equation} 
\alpha_{t}=\frac{\exp \left(u^{\text{tr}} y_t\right)}{\sum_{t=1}^{T} \exp \left(u^{\text{tr}} y_t\right)},
\label{eq:attention_weights}
\end{equation}

where the superscript $tr$ denotes a transpose operation, and $u$ and $y$ stand, respectively, for the attention parameter vector, which is learnable, and the LSTM output. The inner product between $u$ and $y_t$ measures the importance of each $t$-th frame. Then, a softmax transformation is applied in order to normalize the weights, guaranteeing that their sum across all the frames of the utterance sum up to one. Both, the attention parameters and the LSTM outputs, are obtained in the whole training process of the system.

\begin{figure}[t]
	\centering
	\includegraphics[width=0.6\textwidth]{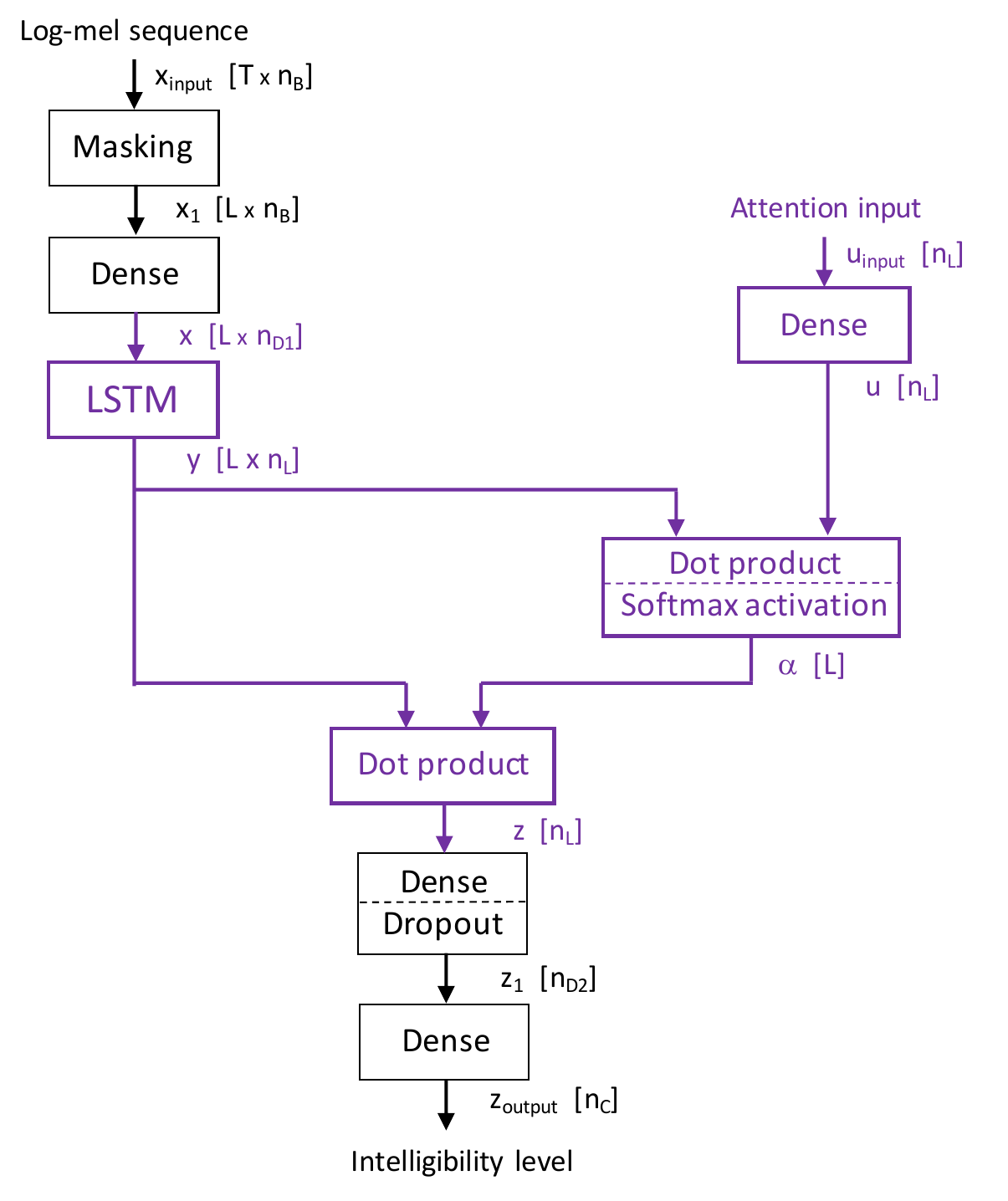}
	\caption{LSTM-based architecture with the attention mechanism. In brackets, the dimension of each variable, where $T$, $n_B$, $L$, $n_{D1}$, $n_L$, $n_{D2}$ and $n_C$, stand for the number of frames of the input signal, the number of mel filters, the length of the LSTM input/output sequence, the number of neurons in the first dense layer, the number of LSTM units, the number of neurons in the second dense layer and the number of classes (intelligibility levels), respectively.}
\label{fig:lstm_attention_architecture}
\end{figure}

The block diagram of the LSTM with Attention system is represented in Figure \ref{fig:lstm_attention_architecture}. As can be observed, the architecture is the same as for the basic LSTM and LSTM with Mean-Pooling systems, except for the layers involved in the attention mechanism (depicted in purple lines).

\section{Experiments}
\label{sec:experiments}

\subsection{Database}
\label{subsec:database}

The dataset used for the experiments is the UA-Speech database \citep{Kim2008}. It consists of audio files with voice of 15 persons (11 men and 4 women) that suffer from dysarthria with different degrees, as well as recordings from 13 healthy control people (9 men and 4 women). The audio files were recorded  at 16 KHz with an array of 7 microphones and contain utterances such as digits, computer commands, simple short words, multisyllable complex words, and the radio alphabet.

The database was annotated in terms of the so-called intelligibility score, by means of medical tests carried out by specialists that listened to the recordings and wrote down the percentage of understood words. The resulting labels have a range between 0 and 100 being 0 completely unintelligible and 100 perfectly intelligible. However, as our task is to predict the intelligibility level \textcolor{black}{into} three categories, \textcolor{black}{these former labels were modified}, in such a way that scores from 0 to 33 correspond to the low intelligibility class (L), from 34 to 66 to the medium intelligibility one (M), and from 67 to 100 to the high intelligibility one (H).

Since the purpose of this work is to develop an automatic speech intelligibility system following a non intrusive approach, the audios from the healthy controls were not used at all. Moreover, only the speech signals recorded on the sixth microphone were considered. In summary, the total number of available files for the experiments is $9,140$.

The experiments have been configured in a speaker-independent way, so training, \textcolor{black}{validation, and test} subsets contain disjoint speakers. The approximate percentages of files for these three partitions are 50\%, \textcolor{black}{15\%, and 35\%}, respectively. \textcolor{black}{This set-up is intended for avoiding} a bias in the results, as the developed systems might learn the speakers' identity or the environmental acoustic conditions instead of their intelligibility level if same speakers are included in training, \textcolor{black}{validation, and test} splits.

\subsection{Preprocessing and feature extraction}

As mentioned in Subsection \ref{subsec:preprocessing}, in several experiments a VAD has been included to remove the silence and other non-speech frames of the recordings. \textcolor{black}{In particular, the implementation of the Sohn's VAD available in the VOICEBOX toolbox \citep{VoiceBox} has been used}. It is worth noting that, as shown in Subsection \ref{subsec:results}, this VAD step was finally omitted when tackling the problem by using the LSTM approach because results suggest that this kind of networks are capable of extracting some useful information from the characteristics of the pauses and the presence of disfluencies or other speech artifacts that are phenomena related to the intelligibility level. 

For the reference system, all the acoustic parameters were computed every $10~ms$ using a Hamming window of $20~ms$ long \textcolor{black}{with} the toolboxes VOICEBOX \citep{VoiceBox} and SRMR \citep{Falk2010}. This last software was utilized for the calculation of the modulation spectrum and the LHMR parameter.

In the case of the MFCC parameterization, $13$ cepstral coefficients and their corresponding first derivatives were computed from an auditory filter bank composed of $40$ triangular mel-scaled filters, yielding to $26$ dimension feature vector.

The average energy of the modulation spectrum was computed using $23$ critical band frequencies, which cover the range in which humans generally articulate the speech, and $8$ modulation filters according to the recommendations in \citep{Falk2012, SarriaPaja2012}. In this case, the final feature vectors consists of $23 \; \text{x} \; 8 = 184$ coefficients.

The Falk's parameterization consists of the six features described in Subsection \ref{subsubsec:falk_features} and were computed following the set-up mentioned in \citep{Falk2012, SarriaPaja2012}.

For the LSTM-based approaches, log-mel spectrograms were used as input to the networks. This feature set consists of $n_B = 32$ log-Mel filterbank energies, computed every $10~ms$ \textcolor{black}{with} a Hamming window of $20~ms$ long and a mel-scaled filterbank composed of $n_B = 32$ filters by using the  Python's package LibROSA \citep{LibROSA}. Mean and standard deviation normalization are applied at utterance-level yielding to a set of normalized log-mel spectrogram sequences $x_{input}$ with $T \times 32$ dimensions, where $T$ is the number of frames of each utterance.

\begin{figure}[t]
	\centering
	\includegraphics[width=0.5\textwidth]{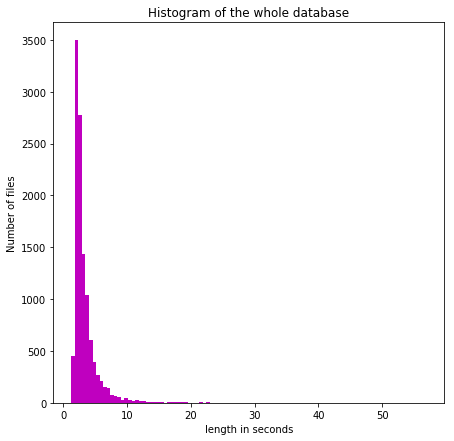}
	\caption{Histogram of the length of the audio files in the database.}
\label{fig:histogram}
\end{figure}

As efficient computation of LSTM networks requires that all the input sequences have the same dimensions, a fixed length of $L = 700$, which corresponds to $7~s$ was established. If the audio recordings were longer than this amount, then the corresponding log-mel spectrograms were cropped, and, otherwise, they were padded with some dummy values. Note that the use of the appropriate Masking layers in the LSTM-based architectures easily allow to ignore these masked values in further computations. The quantity $L = 700$ was decided by looking at the histogram of the duration of the audio signals in the database, which is depicted in Figure \ref{fig:histogram}. As can be seen, more than the $95\%$ of the files have a duration of $7~s$, so only the remaining $5\%$ of the recording were cut. 

\subsection{Classifiers}

The reference system is based on a SVM running on normalized features and it was developed using the MATLAB Statistics and Machine Learning toolbox. The optimal hyperparameters of the model were obtained by means of a Bayesian optimizer with a 5-fold cross validation strategy.

\textcolor{black}{All the LSTM-based systems} were implemented with the Tensorflow \citep{Abadi2015} and Keras \citep{Chollet2015} packages. The specific values of the configuration parameters of the three different architectures depicted in Figures \ref{fig:lstm_architecture} (a) and (b) and Figure \ref{fig:lstm_attention_architecture} can be found in Table \ref{tab:lstm_configuration}. In all cases, the LSTM models were trained using stochastic gradient descent and the Adam method with an initial learning rate of $0.0002$. \textcolor{black}{A batch size of 32 and a maximum number of 40 epochs were considered}. In the indicated layers, a $33 \%$ dropout was set to avoid over-fitting in the training process.

\begin{table}[t]
\begin{center}
\begin{adjustbox}{max width=1.0\textwidth}
\begin{tabular}{|c|c|c|}
\hline 
		 & 			& \\
Parameter & Description & Value \\ 
\hline \hline 
$T$ & No. of frames of the input signal & Variable \\
\hline 
$L$ & No. of frames of the LSTM input/output sequences & 700 \\
\hline 
$n_B$ & No. of mel filters in the log-mel spectrogram computation & 32 \\
\hline 
$n_{D1}$ & No. of neurons in the first dense layer & 32 \\
\hline 
$n_L$ & No. of LSTM units & 128 \\
\hline
$n_{D2}$ & No. of neurons in the second dense layer & 50 \\
\hline 
$n_C$ & No. of classes (intelligibility levels) & 3 \\
\hline 
\end{tabular} 
\end{adjustbox}
\end{center}
\caption{Value of the configuration parameters for the LSTM-based systems.}
\label{tab:lstm_configuration}
\end{table}

In the attention model, the attention parameter vector $u$ has a dimension of $n_L = 128$. All its components were initialized to $1/L$ and then refined during the training stage of the whole system.

\subsection{Results}
\label{subsec:results}

%This Subsection contains the experiments carried out in order to assess the performance of the proposed LSTM-based systems.
\textcolor{black}{The metric used for the assessment of the developed systems is the \emph{accuracy or classification rate per audio recording}, that is defined as the number of correctly classified files divided by the total number of tested files. In the case of the LSTM-based systems, each experiment was run 20 times and therefore, results corresponding to systems where LSTMs are involved are the average accuracy across the 20 subexperiments and the respective standard deviation.}

\textcolor{black}{Table \ref{tab:results_svm} contains the results achieved by the reference system with different parameterizations and set-ups: with VAD (\emph{SVM + VAD}), without VAD (\emph{SVM}), and when a frame weighting scheme is applied to the features (\emph{SVM + Attention Weights}). In this latter case, for comparison purposes, the weights were computed according to Equation \eqref{eq:attention_weights} by using the Attention LSTM architecture depicted in Figure \ref{fig:lstm_attention_architecture}. Table \ref{tab:results_lstm} shows the accuracies obtained with the LSTM-based systems proposed in this work (\emph{Basic LSTM + VAD}, \emph{Basic LSTM}, \emph{LSTM Mean-Pooling}, and \emph{LSTM with Attention-Pooling}).}

\begin{table}[t]
\begin{center}
\begin{adjustbox}{max width=0.75\textwidth}
\begin{tabular}{|c|c|c|}
\hline 
		 & 			& \\
System & Features & Accuracy [\%] \\ 
\hline \hline 
SVM + VAD & MFCC + $\Delta$MFCC & 44.80\% \\
\hline 
SVM + VAD & Modulation Spectrum & 45.25\% \\
\hline 
SVM + VAD & Falk's features & 40.71\% \\
\hline \hline
SVM & MFCC + $\Delta$MFCC & 41.26\% \\
\hline 
SVM & Modulation Spectrum & 47.32\% \\
\hline 
SVM & Falk's features & 55.54\% \\
\hline \hline
\textcolor{black}{SVM + Attention Weights} & \textcolor{black}{MFCC + $\Delta$MFCC}  & \textcolor{black}{44.09\%$\pm$0.42\%} \\
\hline 
\textcolor{black}{SVM + Attention Weights} & \textcolor{black}{Modulation Spectrum} & \textcolor{black}{52.83\%$\pm$0.47\%} \\ 
\hline
\textcolor{black}{SVM + Attention Weights} & \textcolor{black}{Falk's features} & \textcolor{black}{\textbf{61.05\%$\pm$0.36\%}} \\
\hline
\end{tabular} 
\end{adjustbox}
\end{center}
\caption{\textcolor{black}{Classification rates [\%] achieved by the reference system with different features.}}
\label{tab:results_svm}
\end{table}

\begin{table}[t]
\begin{center}
\begin{adjustbox}{max width=0.75\textwidth}
\begin{tabular}{|c|c|c|}
\hline 
		 & 			& \\
System & Features & Accuracy [\%] \\ 
\hline \hline 
Basic LSTM + VAD & Log-mel spectrogram & 63.11\%$\pm$0.39\% \\   
\hline 
Basic LSTM & Log-mel spectrogram & 67.50\%$\pm$0.29\% \\
\hline 
LSTM Mean-Pooling & Log-mel spectrogram & 74.82\%$\pm$0.39\% \\ 
\hline
LSTM Attention-Pooling & Log-mel spectrogram & \textbf{76.97\%$\pm$0.28\%} \\
\hline
\end{tabular} 
\end{adjustbox}
\end{center}
\caption{\textcolor{black}{Classification rates [\%] achieved by the LSTM-based classifiers.}}
\label{tab:results_lstm}
\end{table}

First of all, \textcolor{black}{the suitability of the VAD as preprocessing step for the task of intelligibility level prediction is analyzed}. As can be observed, for both, SVM and LSTM-based systems, the use of a VAD leads to a worse performance in all cases with the exception of the reference system with MFCC parameterization. This \textcolor{black}{suggests} that silence pauses and other speech artifacts (such as stuttering or hesitations) convey relevant information for the systems in order to make better predictions, as they are related to the rhythm, elocution \textcolor{black}{rate, and the} presence of disfluencies that are characteristics of the dysarthric speech \citep{Doyle1997, Falk2012}. For this reason, the remaining experiments were carried out without VAD.

As for the comparison of the reference system with different sets of features, it can be seen that the Falk's parameterization is the best option followed by the modulation spectrum. These results have sense as, in contrast to MFCC which are general purpose parameters for speech tasks, Falk's features were specifically designed for speech intelligibility prediction. In addition, \textcolor{black}{as it was hypothesized} in Subsection \ref{subsubsec:modulation_spectrum}, the modulation spectrum contains some important cues about the comprehensibility of an utterance, although it should be combined with other information sources, as for example, prosody-related parameters. \textcolor{black}{Finally, it is worth noting that the use of attention weights improve the results for all the parameterizations. This behavior suggests that it is important to emphasize the contribution of the more relevant frames to the task independently of the classifier used. Again, in this set-up, Falk's features are the best choice for the SVM-based system.} 

\textcolor{black}{Regarding the LSTM-based systems, on the one hand, \emph{Basic LSTM} outperforms significantly the best reference system (\emph{SVM + Falk's features + Attention weights}), obtaining a relative error reduction of 16.55\%.} This confirms that the temporal modeling performed by LSTM is crucial for addressing the problem of speech intelligibly classification. On the other hand, further improvements can be achieved when applying the Mean-Pooling strategy (\emph{LSTM Mean-Pooling}), and therefore, it seems better not to completely discard LSTM frames.

Moreover, the attention mechanism seems to learn the important parts of the utterances, so it is useful tool for this type of tasks when working with LSTMs. In fact, \emph{LSTM Attention-Pooling} obtains the best performance in comparison to all SVM-based systems and the remaining LSTM models. More specifically, \textcolor{black}{this approach achieves 40.87\% relative error reduction with respect to \emph{SVM + Falk's features + Attention weights}} and 8.54 \% with respect to \emph{LSTM Mean-Pooling}.

\begin{figure}[t]
	\centering
	\includegraphics[width=1.0\textwidth]{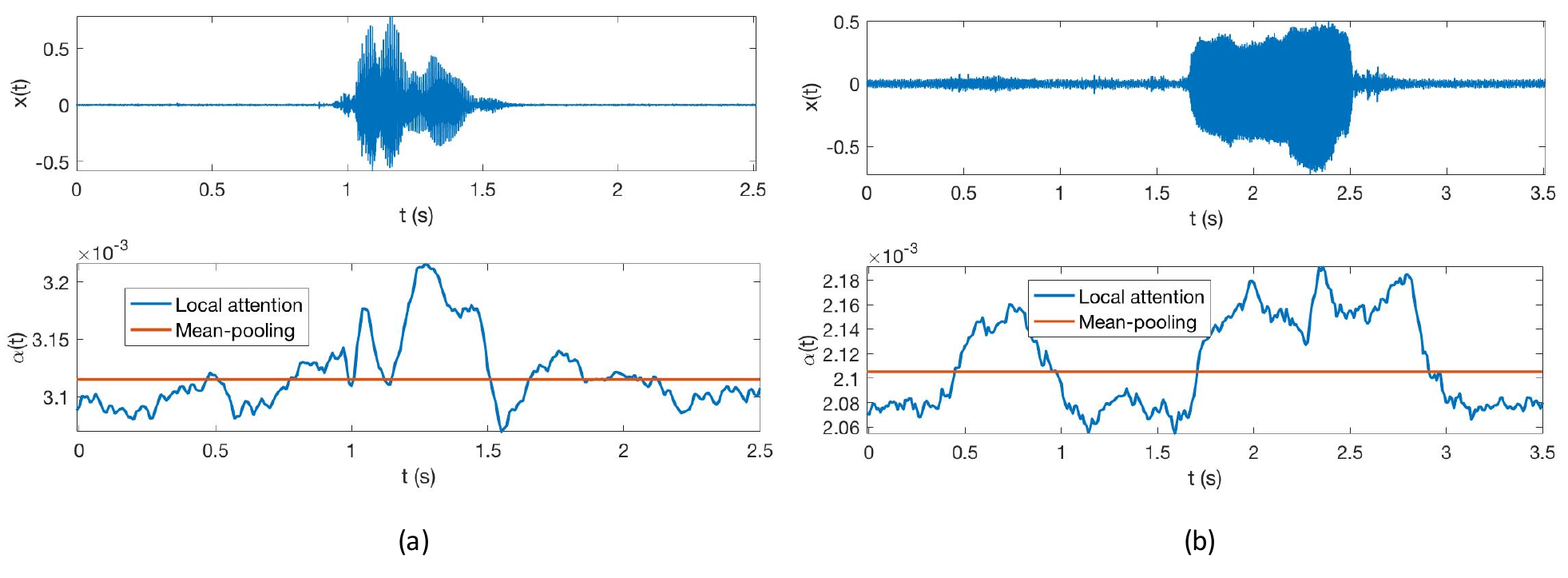}
	\caption{Mean-pooling and Attention weights for an utterance with (a) high intelligibility  and (b) low intelligibility. Top: Waveform. Bottom: Weights corresponding the mean-pooling approach (red line) and to the attention model (black line). Both utterances correspond to the word ``jowls".}
\label{fig:weights}
\end{figure}

Figure \ref{fig:weights} depicts the waveform (top) and its corresponding Mean-Pooling and Attention weights (bottom) for an utterance with (a) high intelligibility and (b) low intelligibility. In contrast to Mean-Pooling weights, which are constant, Attention weights present a large degree of variation. In general terms, larger weights correspond to speech segments of high energy, whereas smaller ones are assigned to silence or low energy frames. Nevertheless, a more in-depth analysis of the example of low intelligibility (see Figure \ref{fig:weights} (b)) allows to draw two interesting observations. Firstly, weights of silence and non-speech frames are small but greater than zero. This fact corroborates the hypothesis that this kind of frames might convey useful information about the utterance intelligibility. Secondly, disfluencies and hesitations are assigned to high weights even when their energy is low (see the segment from $0.5$ to $1~s$), showing the importance of these speech artifacts for the task under consideration. 

\section{Conclusions}
\label{sec:conclusions}

In this work, \textcolor{black}{it has been proposed} an automatic non-intrusive system designed for speech intelligibility level classification. The system is based on LSTM networks and uses log-mel spectrograms as input features. Moreover, a simple attention mechanism has been incorporated into the LSTM-based architecture in order to perform a more adequate modeling of the log-mel spectrogram sequences.

\textcolor{black}{The developed systems have been evaluated} over the UA-Speech database that contains dysarthric speech with different levels of severity. Results have shown that the proposed architectures based on LSTMs and log-mel spectrograms outperform significantly the reference system that is based on SVM and hand-crafted features. In addition, the LSTM system that uses the attention mechanism achieves better results than the Basic LSTM and LSTM with Mean-Pooling alternatives. In particular, the Attention LSTM model achieves \textcolor{black}{a relative error reduction of 40.87\% with respect to the best SVM-based system (where Falk's features with Attention weights are used)} and 8.54\% with respect to LSTM with Mean-Pooling.

Some future lines that have been taken into consideration for the LSTM-based systems include the usage of the modulation spectrum as input feature (instead of log-mel spectrograms or in combination with them), the development of more complex neural network architectures, and the incorporation of auditory saliency features \citep{Kaya2017} for weights computation in the attention mechanism. \textcolor{black}{Finally, a parallel line of work will focus on the study of the classification errors produced by the different systems as a function of several factors, such as utterance duration, pronunciation difficulty, and phoneme similarity and confusability.}

\section*{Acknowledgments}

The work leading to these results has been partly supported by the Spanish Government-MinECo under Project TEC2017-84395-P. The authors wish to acknowledge Dr. Mark Hasegawa-Johnson for making the UA-Speech database available.

%\section*{References}

\bibliography{mybibfile}

\end{document}